\documentclass[aip,jcp,reprint,onecolumn,12pt,amsmath,amssymb,amsfonts,floatfix,letterpaper]{revtex4-1}

\usepackage{graphicx}
\usepackage{braket}
\usepackage{tablefootnote}
\usepackage{color}
\usepackage{enumitem}

\begin{document}

\title{General-order open-shell coupled-cluster method with partial spin adaptation I: formulations \vspace*{0.2cm}}
\author{Cong Wang}
\email{congwang.webmail@gmail.com}
\affiliation{PO Box 26 Okemos, MI, 48805 (USA)}
\affiliation{The Pennsylvania State University; 401A Chemistry Building; University Park, PA 16802 (USA)}

\begin{abstract}

A general-order open-shell coupled-cluster method based on spatial orbitals is formulated. The method is an extension of the partial-spin adaptation (PSA) scheme from Janssen and Schaefer (Theor. Chim. Acta, 79, 1-42, 1991). By increasing the order of excitation operator and spin adaptation, the full configuration interaction (CI) limit is expected to be achieved. In the meanwhile, the advantages of spin-free approach in closed-shell systems: fourth-order termination of the Baker-Campbell-Hausdorff (BCH) expansion and only $HT$-type connected terms are conserved.

\end{abstract}

\maketitle

\section{Introduction}

Coupled-cluster (CC) theory is one of the most successful methods for accurately describing many-electron correlations \cite{bartlett1995coupled, christiansen2006coupled, Bartlett:07,crawford2007introduction,Bartlett:09, bartlett2012coupled,     Lyakh:12,bartlett2024perspective}.
With the help of automatic generation algorithms, any order of excitations in the spin-orbital formalism can be obtained  \cite{Harris:99,Olsen:00,Hirata:00, hirata2003tensor, Kallay:00,kallay2001higher, auer2006automatic, parkhill2010sparse,engels:11,evangelista2022automatic}. Subchemical accuracy  (0.1 kcal mol $^{-1}$) may be achieved at quadruple excitations for weakly correlated systems \cite{martin1999towards,tajti2004heat,yang2014ab}. Since the non-relativistic Hamiltonian is spin-free, it is possible to use the spatial orbitals to reduce a large prefactor of computational cost. Recently, much progress has been made in closed-shell systems \cite{Stanton:14,Stanton:15,matthews2019diagrams}.

The spatial-orbital formulation of the open-shell CC theory is substantially more complex than the closed-shell methods. The most straightforward approach is to adopt the spin-free excitation operators directly. However, it has been realized that this approach would let the BCH expansion not be terminated at the fourth order \cite{Janssen:91,herrmann2022analysis}, and the elimination of disconnected terms is no longer valid\cite{Janssen:91}. Various approaches have been proposed: truncating the BCH expansion \cite{Li:94,Jeziorski:95a,Jeziorski:95b,Li:98,nooijen2001towards, Datta:08,Datta:13}, PSA \cite{Janssen:91,Knowles:93,neogrady1994spin, Neogrady:95,urban1997spin, grotendorst2000modern,knowles2000erratum},
using spin-orbital equations to constrain the amplitudes \cite{piecuch1989orthogonally,Szalay:97,Szalay:00,Heckert:06},  normal-ordering the wave operator \cite{lindgren1978coupled,   mukherjee1979hierarchy, haque1984application, mukherjee1989use,   nooijen1996many, nooijen1996general,  Datta:08, datta2011state, Datta:13, datta2014analytic, datta2015communication, datta2019accurate, gunasekera2024multireference}, performing a CI-type expansion on top of a closed-shell reference state or approximating the cluster expansion \cite{nakatsuji1977cluster, nakatsuji1978cluster_a, nakatsuji1978cluster, nakatsuji1979cluster, nakatsuji1983cluster,jayatilaka1993open}, introducing a bath system \cite{folkestad2023entanglement},
and recently fully spin-adapted open-shell CC method \cite{herrmann2020generation} including its simplifications \cite{herrmann2022analysis,herrmann2022correctly}.

In the present work, we aim to develop a general-order open-shell coupled-cluster method. The target approach should:
\begin{enumerate}
 \item[(i)] employ spatial-orbital;
 \item[(ii)] converge to the full-CI limit;
 \item[(iii)] be potentially useful, namely if the full spin adaptation is over expansive, further approximations will be adopted.
\end{enumerate}

We think the PSA scheme would be adequate for achieving  the aforementioned objectives.
It has the same number of parameters as directly using the spin-free excitation operators.
The merits in closed-shell CC formulation: fourth-commutator termination of the BCH expansion and only $HT$-type connected contractions (besides the terms from projection manifolds), are preserved. The missing components for a complete spin adaptation can be supplemented by modifying the lower-order excitations.

In this article, we shall first review the framework of second quantization with the help of generalized normal ordering \cite{Kutzelnigg:97}. The normal-ordered Hamiltonian is set up accordingly. We then discuss the CC parametrization of an open-shell state and extend the PSA into general order.

Since our approach only includes $HT$-type connected contractions, it would have considerable merit in organizations of equations compared to Herrmann and Hanrath's approach \cite{herrmann2020generation,herrmann2022analysis,herrmann2022correctly}.

\section{Theory}

\subsection{Tensor notation and normal ordering}

We employ the {\it Quantum Chemistry in Fock Space} \cite{Kutzelnigg:97} framework.
Einstein summation convention is assumed for all indices except the spin indices. $IJKL$, $ABCD$, $VWXY$, and $PQRS$ denote closed, external, active, and general spatial orbitals, respectively \cite{Helgaker:00book}.
Quantities with and without tilde symbol, $\tilde{}$, indicate normal ordered or not entries.
In this section, we start our discussion from the spin-orbital basis, then introduce the spin-free quantities by summing up the spin components. $a$ and $\gamma$ stand for operator and density matrix in spin-orbital basis, respectively.
$E$ and $\Gamma$ stand for operator and density matrix in spatial orbital quantities, respectively. For an operator in the PSA scheme, it is possible to introduce mixed spin summed and unsummed entries.
For example, $E^A_I a^{B\alpha}_{V\alpha} := \sum_{\sigma=\alpha,\beta} a^{A\sigma}_{I\sigma} a^{B\alpha}_{V\alpha}  $, where $\alpha$ and $\beta$ are spin indices. We adopt $E^{A B\alpha}_{I V\alpha}$ for brevity  (including $E^{B\alpha}_{V\alpha} := a^{B\alpha}_{V\alpha}$).

The normal-ordered spin-orbital excitation tensors are defined as
\begin{align}
\tilde{a}_{R_1 \sigma_1}^{S_1 \sigma_1} &:= {a}_{R_1 \sigma_1 }^{S_1 \sigma_1} - \gamma_{R_1 \sigma_1 }^{S_1 \sigma_1} \label{no_1}
\\
\tilde{a}_{R_1 \sigma_1 R_2 \sigma_2 }^{S_1 \sigma_1 S_2 \sigma_2 } &:=
{a}_{R_1 \sigma_1 R_2 \sigma_2 }^{S_1 \sigma_1 S_2 \sigma_2 }
- \gamma_{R_1 \sigma_1 }^{S_1 \sigma_1} \tilde{a}^{S_2 \sigma_2 }_{R_2 \sigma_2}
- \gamma^{S_2 \sigma_2  }_{R_2 \sigma_2 } \tilde{a}^{S_1 \sigma_1 } _{R_1 \sigma_1}
+ \gamma_{R_2 \sigma_2}^{S_1 \sigma_1} \tilde{a}^{S_2 \sigma_2}_{R_1 \sigma_1} \nonumber \\
& + \gamma^{S_2 \sigma_2}_{R_1 \sigma_1 } \tilde{a}^{S_1  \sigma_1 }_{R_2 \sigma_2}
- \gamma^{S_1 \sigma_1 S_2 \sigma_2}_{R_1 \sigma_1 R_2 \sigma_2}  \label{no_2} \\
\cdots  \nonumber
\end{align}
where $\sigma_i, i=1,2,3,\cdots $ is the spin index. ${a}_{R_1 \sigma_1  \cdots  R_k \sigma_k }^{S_1\sigma_1  \cdots S_k \sigma_k}  := {a}_{S_1 \sigma_1}^{\dagger} \cdots {a}_{S_k \sigma_k}^{\dagger} $ $ {a}_{R_k \sigma_k} \cdots  {a}_{R_1 \sigma_1}$ denotes a string of creation and annihilation operators with respect to the genuine vacuum, and
\begin{align}
\gamma_{R_1 \sigma_1 \cdots R_k \sigma_k }^{S_1\sigma_1  \cdots S_k \sigma_k }:= \langle \Phi|{a}_{R_1 \sigma_1 \cdots  R_k \sigma_k }^{S_1\sigma_1 \cdots S_k \sigma_k} \ket{\Phi}
\end{align}
denotes the $k$-electron reduced density matrix (RDM).
In the present scope, the wavefunction $\Phi$ is a Slater determinant obtained from a restricted open-shell Hartree-Fock (ROHF) calculation. Normal-ordered operators of higher order may be defined recursively as described in Ref. \cite{Kutzelnigg:97}.

The corresponding spin-free quantities (without spin indices) are defined as:
\begin{align}
{E}_{R_1 \cdots R_k }^{S_1 \cdots S_k} &:= \sum_{\sigma_1, \cdots, \sigma_k\in\{\alpha,\beta\}} {a}_{R_1 \sigma_1\ \cdots\ R_k \sigma_k }^{S_1 \sigma_1\ \cdots\ S_k \sigma_k} \label{spinsum_0} \\
\tilde{E}_{R_1 \cdots R_k }^{S_1 \cdots S_k} &:= \sum_{\sigma_1, \cdots, \sigma_k\in\{\alpha,\beta\}} \tilde{a}_{R_1 \sigma_1\ \cdots\ R_k \sigma_k }^{S_1 \sigma_1\ \cdots\ S_k \sigma_k} \label{spinfree-excitation} \\
\Gamma_{R_1    \cdots R_k  }^{S_1  \cdots S_k  } &:= \sum_{\sigma_1, \cdots, \sigma_k\in\{\alpha,\beta\}}  \gamma_{R_1 \sigma_1  \cdots R_k \sigma_k }^{S_1\sigma_1 \cdots S_k \sigma_k } \label{spinsum}
\end{align}
The many-electron Hamiltonian can hence be written as
\begin{align}
{H} &= E_{\textrm{ref}} +  \sum_{\sigma} f_{Q \sigma}^{P \sigma} \tilde{a}^{Q\sigma}_{P\sigma} + \frac{1}{2} W_{PQ}^{RS} \tilde{E}_{RS}^{PQ} \label{ab} \\
E_{\textrm{ref}} &=  \sum_{\sigma} h_{P \sigma}^{Q \sigma} \gamma^{P \sigma}_{Q \sigma} + \frac{1}{2} \sum_{\sigma \tau} W_{P\sigma Q\tau}^{R\sigma S\tau} \gamma^{P\sigma Q\tau}_{R\sigma S\tau} \\
 &=   h_{P }^{Q } \Gamma^{P }_{Q } + \frac{1}{2} W_{P Q}^{R S} \Gamma^{P Q}_{R S} \\
f^{P \sigma}_{Q \sigma} &= h^{P \sigma}_{Q \sigma} + \sum_{\tau} ( W^{R \tau P\sigma }_{S \tau Q \sigma} - W^{R \tau P \sigma}_{Q \sigma  S\tau } ) \Gamma^{S \tau}_{R \tau} \label{fock}
\end{align}
where $h$ and $W$ are the one and two-electron integrals, respectively. The relations between values in the spatial and the spin-orbital bases are given as
\begin{align}
h_P^Q &= h_{P \alpha}^{Q \alpha} = h_{P \beta}^{Q \beta}, 0 = h_{P \alpha}^{Q \beta} = h_{P \beta}^{Q \alpha}  \\
W_{PQ}^{RS} &= W_{P \alpha Q\alpha}^{R \alpha S\alpha} =  W_{P \alpha Q\beta}^{R \alpha S\beta} =  W_{P \beta Q\alpha}^{R \beta S\alpha}  =  W_{P \beta Q\beta}^{R \beta S\beta}, 0 = W_{P \alpha Q \beta}^{R \alpha S\alpha} = \cdots
\end{align}
$E_{\textrm{ref}}$ is the ROHF energy.

Assuming only the alpha spin orbitals are singly occupied in the high-spin ROHF reference state, the non-zero elements of the one-electron density matrices are
\begin{align}
{\gamma}^{I \alpha}_{I \alpha} &= {\gamma}^{I \beta}_{I \beta} = 1 \\
{\gamma}^{V \alpha}_{V \alpha} &= 1 \qquad (\textrm{no summation over } I,V)
\end{align}
The two-electron density matrix can be obtained by \cite{Kutzelnigg:97}
\begin{align}
\gamma^{P\sigma Q \tau}_{R\sigma S\tau} = \gamma^{P\sigma}_{R\sigma} \gamma^{Q\tau}_{S\tau} - \gamma^{P\sigma}_{S\tau} \gamma^{Q\tau}_{R\sigma}
\end{align}
because the reference wavefunction is a single Slater determinant.

The direct use of the spin-orbital Fock operator in Eq. \eqref{ab} may have some disadvantages: it introduces an asymmetry between $\alpha$ and $\beta$ spin components, and the Brillouin condition, $\bra{\Phi} [{H}, {E}_{I}^A ] \ket{\Phi}$=0, is not transparent.
This asymmetry follows from our definitions of normal ordering Eqs. \eqref{no_1} and \eqref{no_2}. The density matrices of active orbitals are not symmetric with respect to $\alpha$ and $\beta$ components for a high-spin ROHF wavefunction, since the singly occupied orbitals are only in $\alpha$-spin.

As an alternative, it is possible to introduce another set of Fock matrices for the inactive (closed-closed, closed-external, and external-external) orbital blocks:
\begin{align}
f^{P }_{Q } = \frac{ f^{P \alpha}_{Q \alpha } + f^{P \beta}_{Q \beta} }{2}, \bar{f}^{P }_{Q } =  \frac{ f^{P \alpha}_{Q \alpha } - f^{P \beta}_{Q \beta} }{2}, \{P,Q\} \in \{IJKL\cdots\} \cup \{ABCD\cdots\}\label{f_convert}
\end{align}
Hence, Eq. \eqref{ab} becomes
\begin{align}
H &=  E_{\textrm{ref}}  + f_I^J \tilde{E}_J^I + f_A^I \tilde{E}_I^A + f_I^A \tilde{E}_A^I  + f_A^B \tilde{E}_B^A \nonumber \\
&+  \bar{f}_I^J \tilde{\bar{E}}_J^I + \bar{f}_A^I \tilde{\bar{E}}_I^A + \bar{f}_I^A \tilde{\bar{E}}_A^I  + \bar{f}_A^B \tilde{\bar{E}}_B^A \nonumber \\
&+  [f^{\alpha}]_{V}^{A} \tilde{a}_{A\alpha}^{V\alpha} +  [f^{\alpha}]^{V}_{A} \tilde{a}^{A\alpha}_{V\alpha}  +  [f^{\alpha}]_{I}^{V} \tilde{a}_{V\alpha}^{I\alpha} + [f^{\alpha}]^{I}_{V} \tilde{a}^{V\alpha}_{I\alpha}          + [f^{\alpha}]^{W}_{V} \tilde{a}^{V\alpha}_{W\alpha}  \nonumber \\
&+ [f^{\beta}]_{I}^{V} \tilde{a}_{V\beta}^{I\beta} + [f^{\beta}]^{I}_{V}  \tilde{a}^{V\beta}_{I\beta} + [f^{\beta}]_{A}^{V}  \tilde{a}_{V\beta}^{A\beta} + [f^{\beta}]^{A}_{V}  \tilde{a}^{V\beta}_{A\beta} + [f^{\beta}]^{W}_{V}  \tilde{a}^{V\beta}_{W\beta}  \nonumber \\
&+ \frac{1}{2} W_{PQ}^{RS} \tilde{E}^{PQ}_{RS} \label{f2}
\end{align}
where $\bar{E}^{P}_Q := a^{P\alpha}_{Q\alpha} - a^{P\beta}_{Q\beta}  $.
This approach corresponds to the form of the Hamiltonian in the work of Knowles, Hampel, and Werner\cite{Knowles:93}, and it encodes the Brillouin conditions as
\begin{align}
 f_I^A            &=f_A^I=0  \label{Brillouin-1} \\
 [f^{\alpha}]_V^A &=  [f^{\alpha}]_A^V =0  \label{Brillouin -2} \\
  [f^{\beta}]_V^I &=  [f^{\beta}]_I^V =0 \label{Brillouin-3}
\end{align}

\subsection{PSA parametrization of CC theory}
\label{psa_t}

The CC theory parameterizes an electronic eigenstate $\ket{\Psi}$ via an exponential ansatz
\begin{align}
 \ket{\Psi}  = e^{{T}} |\Phi \rangle \label{cc}
\end{align}
In the standard spin-orbital formulation of $N$-th order CC theory, the cluster operator ${T}$ is defined as
\begin{align}
{T}   &:= \sum_{i=1}^N {T}_i \\
{T}_k &:=   \frac{1}{k!} t_{A_1 \sigma_1 \cdots A_k \sigma_k}^{I_1 \sigma_1 \cdots I_k \sigma_k }
{a}^{A_1 \sigma_1 \cdots A_k \sigma_k }_{I_1 \sigma_1 \cdots I_k \sigma_k} \qquad (k = 1,2, \cdots, N) \label{cluster}
\end{align}
where $t_{A_1 \sigma_1 \cdots A_k \sigma_k}^{I_1 \sigma_1 \cdots I_k \sigma_k }$ are the excitation amplitudes to be determined.
Inserting Eq. \eqref{cc} for the state vector $\ket{\Psi}$, the time-independent Schr\"{o}dinger equation ${H} | \Psi \rangle = E | \Psi \rangle $ becomes
\begin{align}
 {H} e^{{T}} | \Phi \rangle = E e^{{T}} | \Phi \rangle \label{Sch}
\end{align}
In the standard  linked-form \cite{Helgaker:00book} CC formulation, the energy and amplitudes are obtained by first left-multiplying Eq. \eqref{Sch} with $e^{-{T}}$
and then projecting the resulting equations into the manifold spanned by $ \langle \mu | = \{ \langle \Phi |, \langle \Phi_{I_1 \sigma_1}^{{A}_1 \sigma_1} | , \cdots \}$:
\begin{align}
\langle \Phi |e^{-{T}}  {H} e^{{T}} | \Phi \rangle_{\mathrm{c}} =& E  \label{cce} \\
\langle \mu_k |e^{-{T}}  {H} e^{{T}} | \Phi \rangle_{\mathrm{c}} =& 0  \qquad(k =1,2,3,\cdots, N) \label{ccr}
\end{align}
where $\langle \mu_0 | := \langle \Phi |$, $\langle \mu_k | := \langle \Phi^{I_1 \sigma_1 \cdots I_k \sigma_k }_{A_1 \sigma_1 \cdots A_k \sigma_k }  | $,  and the subscript $\mathrm{c}$ stands for only connected terms \cite{Bartlett:09}. Eqs. \eqref{cce} and \eqref{ccr} correspond to energy and residual equations of CC theory, respectively.

Since the spin-summed quantities (\ref{spinsum_0}) - (\ref{spinsum}) are singlet tensors, it is natural to refine the cluster operator $T$ in terms of spin-summed quantities
\begin{align}
T_1 &=  t_A^I E_I^A +  t_A^V E_V^A  +t_V^I   E_I^V  \label{t1_spin_free} \\
T_2 &= \frac{1}{2} \left(  t_{AB}^{IJ} E_{IJ}^{AB}   + t_{AV}^{IJ} E_{IJ}^{AV} +   t_{VW}^{IJ} E_{IJ}^{VW} +    t_{AB}^{VI} E_{VI}^{AB}
+  t_{AB}^{VW} E_{VW}^{AB} +  t_{AW}^{VI} E_{VI}^{AW} \right) \label{t2_spin_free} \\
\cdots \nonumber
\label{rocc}
\end{align}
as the spin-adapted approach \cite{Janssen:91,herrmann2020generation,    herrmann2022analysis,      herrmann2022correctly}.

However, in the beginning of development of the spatial-orbital open-shell coupled-cluster methods \cite{Janssen:91,neogrady1994spin,   herrmann2022analysis}, it was noticed that Eqs.  \eqref{t1_spin_free} and  \eqref{t2_spin_free} would lead to a BCH expansion not terminate at fourth order as in the spin-orbital and closed-shell spin-adapted formalism.

To simplify the contraction structure of the BCH expansion, Janssen and Schaefer proposed the PSA scheme \cite{Janssen:91},
in which the active-orbital $V \beta$ component in the annihilation operator and the $V \alpha$ component in the creation operators are deleted.
For instance, $ t_V^I E_I^V = t_V^I  \sum_{\sigma= \alpha,\beta}  a_{I\sigma}^{V\sigma} \approx t_V^I a_{I\beta}^{V\beta} $ and
$  t_A^V E_V^A = t_A^V \sum_{\sigma= \alpha,\beta} a_{V\sigma}^{A\sigma} \approx t_A^V a_{V\alpha}^{A\alpha} $. Hence, the creation
and annihilation operators of the active orbitals have different spin components by construction. 
In the PSA framework, the BCH expansion will be terminated at fourth order, the elimination of disconnected terms is valid, and only $HT$ connected contractions are involved in the derivations.

Extended to higher orders, the cluster operator ${T}^{\textrm{PSA}}$ is thus defined as

\begin{align}
{T}^{\textrm{PSA}[1]} &:= {T}_1^{\textrm{PSA}[1]} + {T}_2^{\textrm{PSA}[1]} + {T}_3^{\textrm{PSA}[1]} + \cdots \\
{T}_1^{\textrm{PSA}[1]} &:= t^I_A {E}^A_I + t^{I}_{Y}  {E}^{V\beta}_{I\beta}  \nonumber \\
&\qquad + t^{V}_{A} {E}^{A\alpha}_{V\alpha}   \label{psa1}   \\
{T}_2^{\textrm{PSA}[1]}  &:=  \frac{1}{2!} ( t^{IJ}_{AB} {E}^{AB}_{IJ} + t^{I J}_{ A V}  {E}^{A V\beta }_{I J\beta }+ t^{I J  }_{V W} {E}^{V\beta W\beta }_{I\beta J\beta}  \nonumber \\
&\qquad + t^{VI}_{AB} {E}^{A\alpha B}_{V\alpha I} + t^{VI}_{AW}  {E}^{A\alpha W\beta }_{V\alpha I\beta}  \nonumber \\
&\qquad  + t^{VW}_{AB} {E}^{A \alpha B \alpha}_{V \alpha  W \alpha} )  \label{psa_t2_level_1}
\end{align}

\begin{align}
{T}_3^{\textrm{PSA}[1]} &:= \frac{1}{3!} ( t^{IJK}_{ABC} {E}^{ABC}_{IJK} + t^{IJK}_{ABV} {E}^{A B V\beta}_{IJK\beta} + t^{IJK}_{AVW} {E}_{I J\beta  K\beta }^{AV\beta W\beta}
+ t^{VWX}_{IJK}  {E}^{I\beta J\beta K\beta }_{V\beta W\beta X\beta}   \nonumber \\
&\qquad + t^{VIJ}_{ABC} {E}_{V\alpha IJ}^{A\alpha B C} +t^{VIJ}_{AWB}  {E}_{V\alpha I\beta J}^{A\alpha W \beta B} + t^{VIJ}_{AWX}  {E}_{V\alpha I\beta J\beta}^{A\alpha W\beta X\beta}   \nonumber \\
&\qquad + t^{VWI}_{ABC} {E}_{V\alpha W\alpha I}^{A\alpha B\alpha C} + t^{VWI}_{ABX}  {E}^{A\alpha B\alpha X\beta}_{V\alpha W\alpha I\beta}   \nonumber \\
&\qquad  + t^{VWX}_{ABC} {E}_{V\alpha W\alpha X\alpha}^{A\alpha B\alpha C\alpha} )  \\
{T}_4^{\textrm{PSA}[1]}  &:=    \frac{1}{4!} (  t^{IJKL}_{ABCD} E_{IJKL}^{ABCD} + t^{IJKL}_{ABCV} E_{IJKL\beta}^{ABCV\beta} + t^{IJKL}_{ABVW} E_{IJK\beta L\beta}^{ABV\beta W\beta}   \nonumber \\
&\qquad + t^{IJKL}_{AVWX} E_{IJ\beta K\beta L\beta}^{A V\beta W\beta X\beta} + t^{IJKL}_{VWXY} E_{I\beta J\beta K \beta L \beta}^{V\beta W\beta X\beta Y\beta}   \nonumber \\
&\qquad + t_{ABCD}^{VIJK} E_{V\alpha IJK}^{A\alpha B C D}
+ t^{VIJK}_{AWBC}   E_{V\alpha I\beta JK}^{A\alpha W\beta BC} + t^{VIJK}_{AWXB}   E_{V\alpha  I\beta J\beta K}^{A\alpha W\beta X\beta B}  \nonumber \\
&\qquad + t^{VIJK}_{AWXY}    E_{V\alpha I\beta J\beta K\beta}^{A\alpha W\beta X\beta Y\beta} \nonumber \\
&\qquad + t^{VWIJ}_{ABCD}  E_{V\alpha W\alpha I J}^{A\alpha B\alpha C D} + t^{VWIJ}_{ABXC} E_{V\alpha W\alpha I\beta J }^{A\alpha B\alpha X\beta C }
+ t^{VWIJ}_{ABXY} E_{V\alpha W\alpha I\beta J\beta}^{A\alpha B\alpha X\beta Y\beta}     \nonumber \\
&\qquad + t^{VWXI}_{ABCD}   E_{V\alpha W\alpha X\alpha I}^{A\alpha B\alpha C\alpha D} + t^{VWXI}_{ABCY} E_{V\alpha W\alpha X\alpha I\beta}^{A\alpha B\alpha C\alpha Y \beta} \nonumber \\
&\qquad +  t^{VWXY}_{ABCD} E_{V\alpha W\alpha X\alpha Y \alpha}^{A\alpha B\alpha C\alpha D \alpha} )  \label{psa2} \\
\cdots \nonumber
\end{align}
The superscript $[1]$ denotes to the level of spin adaptation, i.e. only $T_1 | \Phi \rangle$ is spin-adapted. We shall discuss this point in the following paragraphs.

\begin{figure}
\begin{center}
\includegraphics[width=6cm,angle=0]{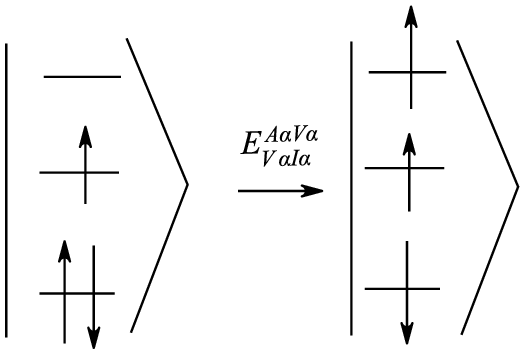}
\caption{A ladder-type excitation produced by the CC expansion. }
\label{fig1}
\end{center}
\end{figure}

One may expect that, expanded to linear order $e^T \approx (1+T)$, the PSA state vector $T | \Phi \rangle$ would be spin-adapted.
However, it has been noticed \cite{Knowles:93,grotendorst2000modern,knowles2000erratum} that in the semi-internal excitation space, $|\Phi^{AW}_{VI}\rangle$, a ladder-type excitation (see Fig. \ref{fig1}) was omitted in Eq. \eqref{psa_t2_level_1}. 

Namely, in excitations of the type
\begin{align}
E^{AW}_{VI}| \Phi \rangle  = ( E_{V \alpha I \beta}^{A \alpha W \beta} - I^W_V E^{A \alpha}_{I \alpha} ) | \Phi \rangle  \label{a3-2}
\end{align} where $I$ is a unit matrix, the $ - I^W_V E^{A \alpha}_{I \alpha} | \Phi \rangle $ term is missing in the PSA[1].

Fortunately, this deficiency can be revised by modifying the single excitation cluster operators as \cite{Knowles:93,grotendorst2000modern,knowles2000erratum}
\begin{align}
t^{I}_{A} E_I^A &\rightarrow  t^{A \alpha }_{I \alpha} E_{I\alpha}^{A\alpha} +    t^{I \beta }_{A \beta} E_{I\beta}^{A\beta} \label{c2_1} \\
t^{I \alpha }_{A \alpha}  &= t^{I }_{A }  - \frac{1}{2} t_{AW}^{VI} I^W_V    \label{c2_2}  \\
t^{I \beta }_{A \beta} &=  t^{I }_{A } \label{c2_3}
\end{align}
without retrieving the fully spin-adapted parametrization of Eqs. \eqref{t1_spin_free} and \eqref{t2_spin_free} and maintain the advantages of PSA (fourth order terminations of the BCH expansion, only connected $HT$ contractions).
We denote this method as $T^{\mathrm{PSA[1|2]}}$. The notation $[1|2]$ stands for generating a spin-adapted state at both the $T_1 |\Phi \rangle$ and $T_2 | \Phi \rangle$ levels. The cluster operator can thus be written as
\begin{align}
T^{\textrm{PSA}[1|2]} &= T^{\textrm{PSA}[1|2]}_1 +  T^{\textrm{PSA}[1]}_2  + T^{\textrm{PSA}[1]}_3 + \cdots  \label{t2-psa-12} \\
T^{\textrm{PSA}[1|2]}_1 &= ( t^{I }_{A }  - \frac{1}{2} t_{AW}^{VI} I^W_V  ) E_{I\alpha}^{A\alpha} + t^{I }_{A } E_{I\beta}^{A\beta}
+ t^{I}_{V}  E^{V\beta}_{I\beta}  + t^{V}_{A} E^{A\alpha}_{V\alpha}\label{t1-psa-12}
\end{align}
 The prefactor $1/2$ in Eq. (\ref{t1-psa-12}) comes from the fact that $1/2$ is in the definitions of $T_2$ terms, Eqs. \eqref{t2_spin_free} and \eqref{psa_t2_level_1}.
At the CCSD level, this cluster operator leads to a CC formulation which is the RHF-RCCSD method by Knowles, Hampel, and Werner \cite{Knowles:93,grotendorst2000modern,knowles2000erratum}.
A more detailed derivation is presented in appendix \ref{appendix_a}.

An objective of the present development is to find higher-order CC expansions capable of systematically converging towards the full-CI solution. To this end, we can consider the spin adaptation for the terms in the fully spin-adapted CC expansion, $e^T$, beyond $T_1$ and $T_2$ levels \cite{paldus1999critical} :
\begin{align}
e^T  | \Phi \rangle  = \left( 1 + T_1  + T_2 + \frac{T_1^2}{2} + T_3 +  \frac{T_1 T_2}{2} +   \frac{T_2 T_1}{2} +   \frac{T_1^3}{3!} + \cdots \right)| \Phi \rangle \label{cc_expan}
\end{align}
Here we do not assume any commutation relation between cluster operators, $T_1, T_2, \cdots$, since $[E_V^A, E_I^W] \neq 0$, if $V=W$ \cite{Helgaker:00book}.

Based on these considerations, in Table \ref{tab1} we provide a list of the modifications of the spin-adapted amplitudes in the CC expansion which are required to reach a certain level of spin adaptation. A derivation can be found in appendix \ref{appendix_a}. The term $ t_{A\alpha B}^{I\alpha J} \mathrel{{+}{=}} - \frac{2}{3!}t_{AWB}^{VIJ}I_{V}^{W}$ in the $T_3$ spin-adaptation corresponds to the "pseudotriple" correction in the approach of Szalay and Gauss \cite{Szalay:00}. In addition, the symmetry properties of the amplitude tensors $t$ is discussed in appendix \ref{appendix_b}.

We expect this type of modifying inactive excitations, e.g., $I\alpha$ to $A\alpha$, could work at general order of CC expansion and spin adaptation, since the contracted active orbital indices are paired with inactive indices (in $E_{V\alpha I\alpha }^{A\alpha V\alpha}$, lower $V$ and upper $V$ are paired with $A$ and $I$, respectively).

\begin{table}[htbp]
  \caption{List of spin-adapted modifications for PSA-CC expansion. $\mathrel{{+}{=}}$ stands for additional terms besides spatial amplitudes, e.g.,  $t_{A \alpha}^{I \alpha} \mathrel{{+}{=}}  - \frac{1}{2} t_{AW}^{VI} I^W_V$
  means $ t_{A \alpha}^{I \alpha} = t_{A  }^{I  } - \frac{1}{2} t_{AW}^{VI} I^W_V $  and $t_{A \beta}^{I \beta} = t_{A}^{I }$ in Eqs. \eqref{c2_2} and \eqref{c2_3}.
  The derivations of these modifications are described in appendix \ref{appendix_a}.
   }
    \begin{tabular}{ll} \hline
    $T_2$ & $t_{A \alpha}^{I \alpha} \mathrel{{+}{=}}  - \frac{1}{2} t_{AW}^{VI} I^W_V$ \\
    $T_1^2$ & $t_{A \alpha}^{I \alpha} \mathrel{{+}{=}}  -  t_{A}^{V} t_{W}^{I} I^W_V$ \\
    $T_3$ & $ t_{A\alpha B}^{I\alpha J} \mathrel{{+}{=}} - \frac{2}{3!}t_{AWB}^{VIJ}I_{V}^{W} $  \\
    & $t_{A\alpha V}^{I\alpha J}  \mathrel{{+}{=}} - \frac{4}{3!}t_{AWV}^{XIJ}I_{X}^{W}$  \\
    & $ t_{AB\alpha}^{VI\alpha}  \mathrel{{+}{=}} - \frac{4}{3!}t_{ABX}^{VWI}I_{W}^{X} $  \\
    $T_1 T_2$   & $ t_{A\alpha B}^{I\alpha J} \mathrel{{+}{=}} -  t_A^V t_{WB}^{IJ}I_{V}^{W} -  t_W^I t_{AB}^{VJ}I_{V}^{W} -  t_B^J t_{AW}^{VI}I_{V}^{W}   $  \\
    & $t_{A\alpha V}^{I\alpha J}  \mathrel{{+}{=}} - 2 t_{A}^X t_{WV}^{IJ}I_{X}^{W}  - 2 t_{W}^I t_{AV}^{XJ}I_{X}^{W}  - 2 t_{V}^J t_{AW}^{XI} I_{X}^{W}  $  \\
    & $ t_{AB\alpha}^{VI\alpha}  \mathrel{{+}{=}} - 2 t_{A}^V t_{BX}^{WI}I_{W}^{X} - 2 t_{B}^W t_{AX}^{VI}I_{W}^{X} - 2 t_{X}^I t_{AB}^{VW}I_{W}^{X}   $   \\
    $T_1^3$  & $ t_{A\alpha B}^{I\alpha J} \mathrel{{+}{=}} - 2 t_{A}^V t_{W}^I t_{B}^{J}I_{V}^{W} $   \\
      &  $t_{A\alpha V}^{I\alpha J}  \mathrel{{+}{=}} - 4 t_{A}^X t_{W}^I t_{V}^{J} I_{X}^{W}$   \\
      &  $ t_{AB\alpha}^{VI\alpha}  \mathrel{{+}{=}} - 4 t_{A}^V t_{B}^W t_{X}^{I}I_{W}^{X} $   \\
      \hline
    \end{tabular}
    \label{tab1}
\end{table}

It may be worthwhile to notice, for the ground state of the lithium atom, if one starts from the fully spin-adapted approach, $E_J^Y E_X^B E_I^W E_V^A  \ket{\Phi} \neq 0$. This implies further spin adaptation beyond Table \ref{tab1} is necessary to obtain the full-CI results.

The aforementioned extensions of the PSA scheme preserves the merits of the closed-shell CC method: fourth order termination of the BCH expansion and only connected $HT$ contractions appear in working equations.  The CC equations, Eqs. \eqref{cce} and  \eqref{ccr}, can be simplified to
\begin{align}
\langle \Phi | {H} e^{{T}^{\mathrm{PSA}}  } | \Phi \rangle_{\mathrm{c}} =& E  \label{cce_c} \\
\langle \Phi | E_{A\cdots}^{I\cdots} {H} e^{{T}^{\mathrm{PSA}} } | \Phi \rangle_{\mathrm{c}} =&   0  \label{ccr_c}
\end{align}
where $\langle \Phi | E_{A\cdots}^{I\cdots}  $ is the excitation manifold $\langle \mu_k^{\textrm{PSA}} |$, that corresponds to the $\langle \mu_k|$ in Eq.  \eqref{ccr}. The specific choice of the excitation operator in $\langle \Phi | E_{A\cdots}^{I\cdots}  $  and  $\langle \mu_k^{\textrm{PSA}} |$ will be discussed in the next subsection.

\subsection{Excitation manifolds}

The cluster operator $T^{\textrm{PSA}}$ encodes the levels of spin adaptations.  The projection functions $\langle  \Phi | E_{A\cdots}^{I\cdots}$ in Eq. \eqref{ccr_c} will have similar forms of spin adaptations. For example, the excitation manifolds,
\begin{align}
\langle \mu_2^{\textrm{PSA}[1]} | &= \langle \Phi | E^{V \alpha I\beta}_{A\alpha W\beta} \label{projection_1} \\
\langle \mu_2^{\textrm{PSA}[1|2]} | &= \langle \Phi | ( E^{V \alpha I\beta}_{A\alpha W\beta} - I_W^V E_{A\alpha}^{I \alpha}) \label{projection_2}
\end{align}
may be used in CCSD calculations. In Eqs. \eqref{projection_1} and \eqref{projection_2}, superscripts $[1]$ and $[1|2]$ stand for the level of spin adaptations, similar to the excitation operators in Eqs. \eqref{t2-psa-12} and \eqref{t1-psa-12}. Expressions up to CCSDTQ are presented in Tables \ref{tab2} and \ref{tab3}.

As pointed out by Knowles, Hampel, and Werner \cite{knowles2000erratum}, if the CC state vector $e^T |\Phi \rangle$ is a spin eigenstate, the level of spin adaptation in the excitation manifolds, $\langle \mu_k^{\mathrm{PSA}} |$, will not affect the energy. Since the PSA wavefunction is not a spin eigenfunction unless the full spin adaptation is adopted, the spin adaptation in the excitation manifolds will have influence on the energy \cite{knowles2000erratum}. Therefore, we denote PSA-T$[1|2]$R$[1|2]$ for $T^{\textrm{PSA}[1|2]}$  \eqref{t2-psa-12} and $\langle \mu_2^{\textrm{PSA}[1|2]} |$ \eqref{projection_2} used in Eqs. \eqref{cce_c} and \eqref{ccr_c}. Similarly,  PSA-T$[1|2|3]$R$[1|2|3]$ stands for $T^{\textrm{PSA}[1|2|3]}$ and $\langle \mu_3^{\textrm{PSA}[1|2|3]} |$.  $T^{\textrm{PSA}[1|2|3]}$ means $T_2$ and $T_3$ spin adaptations in Table \ref{tab1} are adopted, besides $T_1$ spin adaptation is automatically satisfied. $\langle \mu_3^{\textrm{PSA}[1|2|3]} |$ is included in Table \ref{tab2}.

\begin{table}[htbp]
  \caption{List of excitation manifolds used in the present work. $\mu_k,  \qquad k=1,2,3$ stands for $k$-th excitation. They are used to solve the residual CC equations Eq. \eqref{ccr_c}. } 
    \begin{tabular}{ll} \hline
    $\mu_1^{\textrm{PSA}[1]}$ & $\langle \Phi_I^A | = \langle \Phi | E_A^I  $ \\
          & $\langle \Phi_I^V | = \langle \Phi | E_{V\beta}^{I\beta}$ \\
          & $\langle \Phi_V^A | = \langle \Phi | E_{A\alpha}^{V\alpha}$ \\
    $\mu_2^{\textrm{PSA}[1]}$ & $\langle \Phi_{IJ}^{AB} | = \langle \Phi | E_{AB}^{IJ} $ \\
          & $\langle \Phi_{IJ}^{AV}| = \langle \Phi | E_{AV\beta}^{I J\beta}$ \\
          & $\langle \Phi_{IV}^{AB}| = \langle \Phi | E_{AB\alpha}^{I V \alpha }$  \\
          & $\langle \Phi_{IJ}^{VW}| = \langle \Phi | E_{V \beta W\beta}^{I\beta J \beta}$ \\
          & $\langle \Phi_{VI}^{AW}| = \langle \Phi | E_{A \alpha W \beta}^{V \alpha I \beta } $ \\
          & $\langle \Phi_{VW}^{AB}| = \langle \Phi | E_{A \alpha B \alpha}^{V \alpha  W \alpha}$ \\
    $\mu_2^{\textrm{PSA}[1|2]}$ & The same as $\mu_2^{\textrm{PSA}[1]}$ except          \\
                & $\langle \Phi_{VI}^{AW}| = \langle \Phi | ( E_{A \alpha W \beta}^{V \alpha I \beta }  - E_{A\alpha}^{I\alpha} I_W^V )$ \\
    $\mu_3^{\textrm{PSA}[1]}$ & $\langle \Phi_{IJK}^{ABC}| =  \langle \Phi | E_{ABC}^{I J K}$ \\
          & $\langle \Phi_{IJK}^{ABV}| = \langle \Phi | E^{IJK\beta}_{A B V\beta}$  \\
          & $\langle \Phi_{VIJ}^{ABC}| = \langle \Phi | E^{V\alpha I J}_{A\alpha B C} $   \\
          & $\langle \Phi_{IJK}^{AVW}| = \langle \Phi | E^{I J \beta K \beta}_{A V\beta W\beta} $ \\
          & $\langle \Phi_{VIJ}^{AWB}| = \langle \Phi |  E_{A\alpha W\beta B}^{V \alpha I\beta J}     $ \\
          & $\langle \Phi_{VWI}^{ABC}| = \langle \Phi | E^{V\alpha W\alpha I}_{A\alpha B\alpha C} $  \\
          & $\langle \Phi_{IJK}^{VWX}| =  \langle \Phi| E^{I\beta J \beta K \beta}_{V\beta W\beta X\beta} $ \\
          & $\langle \Phi_{VIJ}^{AWX}| =  \langle \Phi |  E_{A\alpha W \beta X\beta}^{V \alpha I\beta J\beta}  $ \\
          & $\langle \Phi_{VWI}^{ABX}| = \langle \Phi |  E_{A\alpha B\alpha X \beta}^{V \alpha W\alpha I\beta} $ \\
          & $\langle \Phi_{VWX}^{ABC}| = \langle \Phi | E^{V\alpha W\alpha X\alpha}_{A\alpha B\alpha C\alpha} $      \\
    $\mu_3^{\textrm{PSA}[1|2]}$        & The same as  $\mu_3^{\textrm{PSA}[1]}$           \\
    $\mu_3^{\textrm{PSA}[1|2|3]}$        & The same as  $\mu_3^{\textrm{PSA}[1]}$     except          \\
                       &  $\langle \Phi_{VIJ}^{AWB}| = \langle \Phi | (  E_{A\alpha W\beta B}^{V \alpha I\beta J} - I_W^V E_{A\alpha B}^{I\alpha J}       ) $ \\
                       & $\langle \Phi_{VIJ}^{AWX}| =  \langle \Phi | (  E_{A\alpha W\beta X\beta}^{V \alpha I\beta J\beta} - I_W^V E_{A\alpha X\beta}^{I\alpha J\beta} -   I_X^V E_{A\alpha W\beta}^{J\alpha I\beta}    )     $ \\
                       & $\langle \Phi_{VWI}^{ABX} |= \langle \Phi | ( E_{A\alpha B \alpha X \beta}^{V \alpha W\alpha I\beta} - I_X^W E_{A\alpha B\alpha}^{V\alpha I\alpha}  -  I_X^V E_{A\alpha B\alpha}^{I\alpha W\alpha}   )  $  \\
      \hline
    \end{tabular}
    \label{tab2}
\end{table}

\begin{table}[htbp]
  \caption{List of the excitation manifolds. The convention of notations in Table \ref{tab2} up to $k=4$ is adopted. }
    \begin{tabular}{ll} \hline
$\mu_4^{\textrm{PSA}[1]}$ & $\langle \Phi_{IJKL}^{ABCD} | =  \langle \Phi | E^{IJKL}_{ABCD}$ \\
& $\langle \Phi_{IJKL}^{ABCV} | =  \langle \Phi | E^{I JKL\beta}_{ABCV\beta}$ \\
& $\langle \Phi_{VIJK}^{ABCD}| =  \langle \Phi | E^{V\alpha IJK}_{A\alpha BCD}$ \\
& $\langle \Phi_{IJKL}^{ABVW}| = \langle \Phi | E^{I JK\beta L\beta}_{ABV\beta W\beta}$ \\
& $\langle \Phi_{VIJK}^{AWBC}| = \langle \Phi | E^{ V\alpha I\beta J K }_{A\alpha W\beta B C} $ \\
&
 $\langle \Phi_{VWIJ}^{ABCD} |= \langle \Phi | E^{V\alpha W\alpha IJ}_{A\alpha B\alpha CD}$ \\
& $\langle \Phi_{IJKL}^{AVWX}| = \langle \Phi | E^{I J\beta K\beta L\beta}_{AV\beta W\beta X\beta}$ \\
& $\langle \Phi_{VIJK}^{AWXB}| = \langle \Phi | E_{A\alpha W\beta X\beta B}^{V\alpha I\beta J\beta K} $ \\
& $\langle \Phi_{VWIJ}^{ABXC}| = \langle \Phi | E_{V\alpha W\alpha I\beta J}^{A\alpha B\alpha X\beta C}  $ \\
& $\langle \Phi_{VWXI}^{ABCD}| = \langle \Phi | E_{A\alpha B\alpha C\alpha D}^{V\alpha W\alpha X\alpha I}$ \\
& $\langle \Phi_{IJKL}^{VXWY}| = \langle \Phi | E_{V\beta W\beta X\beta Y\beta}^{I\beta J\beta K\beta L\beta }  $ \\
& $\langle \Phi_{VIJK}^{AWXY}| = \langle \Phi |  E_{A\alpha W\beta X\beta Y\beta}^{V\alpha I\beta J\beta K\beta} $  \\
& $\langle \Phi_{VWIJ}^{ABXY}| = \langle \Phi |  E_{A\alpha B\alpha X\beta Y\beta}^{V\alpha W\alpha I\beta J\beta}$ \\
& $\langle \Phi_{VWXI}^{ABCY}| = \langle \Phi | E_{A\alpha B\alpha C\alpha Y\beta}^{V\alpha W\alpha X\alpha I\beta}  $\\
& $\langle \Phi_{VWXY}^{ABCD}| = \langle \Phi | E^{A\alpha B\alpha C\alpha D\alpha }_{V\alpha W\alpha X\alpha Y\alpha }$ \\
$\mu_4^{\textrm{PSA}[1|2]}$ & The same as  $\mu_4^{\textrm{PSA}[1]}$            \\
$\mu_4^{\textrm{PSA}[1|2|3]}$ & The same as  $\mu_4^{\textrm{PSA}[1]}$            \\
$\mu_4^{\textrm{PSA}[1|2|3|4]}$ & The same as  $\mu_4^{\textrm{PSA}[1]}$  except            \\
& $\langle \Phi_{VIJK}^{AWBC}| =  \langle \Phi | ( E^{ V\alpha I\beta J K }_{A\alpha W\beta B C} - I^V_W E^{I\alpha JK}_{A\alpha BC} ) $, \\
& $\langle \Phi_{VIJK}^{AWXB}| =  \langle \Phi | (  E_{A \alpha W \beta X \beta B}^{V\alpha I\beta J\beta K} - I_W^V E_{A\alpha X\beta B}^{I\alpha J\beta K} - I_X^V E_{A\alpha W \beta B }^{J\alpha I\beta K}  ) $ \\
& $\langle \Phi_{VWIJ}^{ABXC}| =  \langle \Phi | ( E_{A\alpha B\alpha X\beta C}^{V\alpha W\beta I\beta  J } - I_X^W E_{A\alpha B\alpha C}^{V\alpha I\alpha J} - I_X^V E_{A\alpha B\alpha C}^{I \alpha W \alpha J} ) $ \\
& $\langle \Phi_{VIJK}^{AWXY}| =  \langle \Phi | ( E_{A \alpha W \beta X \beta M\beta}^{V\alpha I\beta J\beta K\beta}
- I_W^V E_{A\alpha X\beta Y\beta}^{I\alpha J\beta K\beta}
- I_X^V E_{A\alpha W \beta Y \beta}^{J\alpha I\beta K\beta}
- I_Y^V E_{A\alpha W \beta X \beta}^{K\alpha I\beta J\beta} ) $ \\
& $\langle \Phi_{VWXI}^{ABCY} |= \langle \Phi | ( E^{V\alpha W\alpha X\alpha I\beta}_{A\alpha B\alpha C\alpha Y\beta}
- I_Y^X E_{A\alpha B\alpha C\alpha }^{V\alpha W\alpha I\alpha } - I_Y^W E_{A\alpha B\alpha C\alpha}^{V\alpha I\alpha X\alpha} - I_Y^V E_{A\alpha B\alpha C\alpha}^{I\alpha W\alpha X\alpha} ) $ \\
& $ \langle \Phi_{VWIJ}^{ABXY}| =  \langle \Phi | ( E^{V\alpha W\alpha I\beta J\beta}_{A\alpha B\alpha X\beta Y\beta}
-  I_X^W E_{A \alpha B\alpha Y\beta }^{V\alpha I\alpha J\beta }
- I_Y^W E_{A\alpha B\alpha X\beta}^{V\alpha J\alpha I\beta}$  \\
& $- I_X^V E_{A\alpha B\alpha Y\beta}^{I\alpha W\alpha J\beta} - I_Y^V E_{A\alpha B\alpha X\beta}^{J\alpha W\alpha I\beta}
+ I_X^V I_Y^W E_{A\alpha B\alpha}^{I\alpha J\alpha}
- I_X^W I_Y^V E_{A\alpha B\alpha}^{I\alpha J\alpha}  ) $ \\
      \hline
    \end{tabular}
    \label{tab3}
\end{table}

\subsection{Evaluations of matrix elements}

To evaluate the matrix elements that include creation and annihilation operators, we employ a straight-forward extension of the  direct  evaluation  of  coupling  coefficients (DECC) scheme of the previous work \cite{wang2018simple}. 

For example, the expectation term in
\begin{align}
\frac{1}{2} \langle   \Phi |  E^{IJ}_{AB}  \tilde{E}^{r\alpha}_{s\alpha}  E^{ab}_{ij} | \Phi \rangle f^{s\alpha}_{r\alpha} t^{ij}_{ab} \label{DECC_os}
\end{align}
may be written by DECC as
\begin{align}
&\frac{1}{2} \langle   \Phi |  E^{IJ}_{AB}  \tilde{E}^{r\alpha}_{s\alpha}  E^{ab}_{ij} | \Phi \rangle = \sum_{\pi\in S_6} C(\pi) \delta^{U_1}_{L_{\pi(1)}}\delta^{U_2}_{L_{\pi(2)}}\ldots\delta^{U_6}_{L_{\pi(6)}}\label{DECC_os2}
\end{align}
Here $\pi$ is a permutation and the factor $C(\pi)$ is determined by
\begin{align}
C(\pi) &= \mathcal{T} \cdot (-1)^{n_\textrm{hole-contractions}} \cdot (-O_{\textrm{cs}})^{n^{\textrm{cs}}_\textrm{cycles-in-$\pi$}} (-O_{{\textrm{os-ss}}})^{n^{{\textrm{os-ss}}}_\textrm{cycles-in-$\pi$}} \nonumber \\
& \qquad (-O_{{\textrm{os-os}}})^{n^{{\textrm{os-os}}}_\textrm{cycles-in-$\pi$}}  \label{eq:CcPrefactor}
\end{align}
where $U$ and $L$ denote to upper and lower indices, respectively, of the matrix element to be evaluated. Eq. \eqref{DECC_os2} shares the general structure of DECC in a closed-shell system. The only difference is the term $O$ in the prefactor $C(\pi)$:
\begin{equation}
 \begin{cases}
  O_{\textrm{cs}} =  2, & \text{All indices in the cycle are inactive}.                                                                     \\
  O_{{\textrm{os-ss}}} = 1, & \text{At least one index in the cycle is active}  \\
       & \text{and all active indices in the cycle have the same spin component}. \\
  O_{{\textrm{os-os}}} = 0, & \text{Both alpha- and beta-spin active indices in the cycle.  }      \nonumber
  \end{cases}
\end{equation}
For $n^{{\textrm{os-os}}}_\textrm{cycles-in-$\pi$} = 0$, we set $(-O_{{\textrm{os-os}}})^{n^{{\textrm{os-os}}}_\textrm{cycles-in-$\pi$}} = 1$.

For example, the permutation $\pi=[4,5,2,1,3]$ in Eq. \eqref{DECC_os2}
\begin{itemize}
   \item Aligns $U=[I,J,r,a,b]$ to $\pi(L)=[i,j,B,A,s]$.
   \item Has two hole contractions ($i$ to $I$ and $j$ to $J$) and two cycles
    ($[4,1]$ and $[3,5,2]$), and therefore a prefactor of $(-1)^2 (-1)^1 (-2)^1=2$.
   \item And thus generates the residual contribution
      \begin{align}
            \frac{1}{2} (2\,\delta^I_i \delta^J_j \delta^r_B \delta^a_A \delta^b_s ) f^s_r t^{ij}_{ab} =  f^{b\alpha}_{B\alpha} t^{IJ}_{Ab}
      \end{align}
\end{itemize}

\section{Summary}

In the present work, we have formulated a general-order open-shell CC method based on the PSA scheme. Progress has been made towards the following objectives: (i) using spatial orbitals; (ii) systematically approaching the full-CI limit; (iii) computational simplicity: preserving the merit of the closed-shell CC method, i.e., fourth-order commutator termination of the BCH expansion and only $HT$ connected contractions. Numerical results are planned to be reported in a subsequent work.

\section*{Acknowledgment}

This work is supported by a starting grant of Pennsylvania State University when the author was in the group of Professor Knizia. The author is indebted for  Professor Knizia for initializing  the project,  the discussions (e.g., noticing $f^{\alpha}$ and $f^{\beta}$ have different orbital energies and remarks on normal-ordered wave operators), and the comments and remarks on the manuscript.

\appendix

\numberwithin{equation}{section}

\section{Derivations for General-Order Spin Adaptations}\label{appendix_a}
In this appendix, we present derivations of the general-order spin adaptations for the amplitudes in Tables \ref{tab1} - \ref{tab3}  by a few examples. We first consider the results of fully spin adapted operators acting on an ROHF reference state. We then use the PSA formulation to incorporate the spin adapted states.

We consider the spin-free excitation operators $T_1$ and $T_2$ in Eqs. \eqref{t1_spin_free} and \eqref{t2_spin_free} acting on an ROHF state $|\Phi\rangle$: 
\begin{align}
T_1 | \Phi \rangle &=  t^I_A E^A_I | \Phi \rangle  + t^I_V E^{V \beta}_{I \beta} | \Phi \rangle + t^V_A E^{A\alpha}_{V\alpha} | \Phi \rangle \label{a1} \\
T_2 | \Phi \rangle &= \frac{1}{2} t^{IJ}_{AB} E^{A B}_{IJ} | \Phi \rangle + \frac{1}{2} t^{IJ}_{AV} E^{AV\beta}_{I J\beta} | \Phi \rangle +   \frac{1}{2}  t^{IJ}_{VW} E^{V \beta W \beta}_{I \beta J \beta}  | \Phi \rangle \nonumber \\
 &+ \frac{1}{2} t_{AB}^{VI} E^{A B \alpha}_{I  V \alpha} | \Phi \rangle + \frac{1}{2} t_{AW}^{VI} E^{AW}_{VI} | \Phi \rangle
 + \frac{1}{2} t_{AB}^{VW} E^{A\alpha B \alpha}_{V\alpha W \alpha} | \Phi \rangle \label{a2}
\end{align}
The RHS of Eq. (\ref{a1}) includes the same terms as  $T_1^{\textrm{PSA[1]}} | \Phi \rangle$.
The RHS of Eq. (\ref{a2}) includes the same terms as  $T_2^{\textrm{PSA[1]}} | \Phi \rangle$ except the fifth term. The fifth term in the RHS in Eq. \eqref{a2}  may be written as
\begin{align}
\frac{1}{2} t_{AW}^{VI}   E^{AW}_{VI}| \Phi \rangle
&= \frac{1}{2} t_{AW}^{VI}  E_{V \alpha I \beta}^{A \alpha W \beta} | \Phi  \rangle - \frac{1}{2} t_{AW}^{VI} I^W_V E^{A \alpha}_{I \alpha} | \Phi \rangle  \label{a3}
\end{align}
The first term in the RHS of Eq. \eqref{a3} appears in $T_2^{\textrm{PSA}[1]}$. The second term, $- I^W_V E^{A \alpha}_{I \alpha}$ ,
implies that the spin-adapted double excitation can be migrated to the single excitation by modifying the specific spin component in Eqs. \eqref{c2_1} - \eqref{c2_3}.
The spin-adapted excitation $( E_{V \alpha I \beta}^{A \alpha W \beta} - I^W_V E^{A \alpha}_{I \alpha} ) | \Phi \rangle$ corresponds to the excitation manifold in Table \ref{tab2}.

Notice the choice of modifying single excitations is not unique, since we can reach the same state by any linear combination of
$( E_{I\alpha}^{A\alpha} + E_{I\beta}^{A\beta} ) | \Phi \rangle $ and $( E_{V\alpha}^{A\alpha} E_{I\beta}^{V\beta} - \frac{1}{2} E^{A\alpha}_{I\alpha} + \frac{1}{2} E^{A\beta}_{I\beta} )| \Phi \rangle $
\cite{Knowles:93} .

Similarly the fully spin-adapted (spin-free) $T_1 T_1$ term on the ROHF reference state $|\Phi \rangle$ is
\begin{align}
\frac{1}{2} T_1 T_1 |\Phi \rangle &= \frac{1}{2} ( E^A_V E^W_I + E^W_I  E^A_V ) t_A^V t_W^I |\Phi \rangle + \cdots  \label{t1t1} \\
&= \frac{1}{2} ( E^{A\alpha}_{V\alpha} E^{W\beta}_{I\beta} + E^{A\beta}_{V\beta} E^{W\beta}_{I\beta}  + E^{Y\alpha}_{I\alpha}  E^{A\alpha}_{V\alpha}  + E^{W\beta}_{I\beta}  E^{A\alpha}_{V\alpha} ) t_A^V t_W^I |\Phi \rangle + \cdots \\
&= \frac{1}{2} ( - E^{A\alpha}_{I\alpha} + E^{A\beta}_{I\beta} ) t_A^V t_W^I |\Phi \rangle +   E^{A\alpha}_{V\alpha} E^{W\beta}_{I\beta}t_A^V t_W^I  |\Phi \rangle  +  \cdots  \label{t1t1-result}
\end{align}
On the RHS of Eq. (\ref{t1t1}), the ellipsis $\cdots$ includes other excitations, besides the semi-internal one, $|\Phi_{VI}^{AW}\rangle$.
Adding $-1/2 ( E_{I\alpha}^{A\alpha} + E_{I\beta}^{A\beta} ) | \Phi \rangle $ on the RHS of Eq. (\ref{t1t1-result}) leads to the expression in the row of $T_1^2$ in Table \ref{tab1}. The non-commuting relation of $[E_V^A, E_I^W] \neq 0$ does not complicate the derivation. In Eq. (\ref{t1t1-result}), $ - E^{A\alpha}_{I\alpha}$   and $E^{A\beta}_{I\beta}$ come from  $E^{A\beta}_{V\beta} E^{W\beta}_{I\beta}$ and $E^{Y\alpha}_{I\alpha}  E^{A\alpha}_{V\alpha}$, respectively. Nevertheless, utilizing the non-uniqueness of spin-adaptation as adding $E_I^A$, $ - E^{A\alpha}_{I\alpha}$   and $E^{A\beta}_{I\beta}$  can be converted to a single term.

Expressions for spin adaptations with triple excitations in Tables \ref{tab1} - \ref{tab3} can be obtained in analogy with the process for double excitations. The effect of fully spin-adapted (spin-free) $T_3$ on the reference ROHF state $|\Phi \rangle$
\begin{align}
 \frac{1}{3!} t_{AWB}^{VIJ} E^{AWB}_{VIJ}  |\Phi \rangle  &=  \frac{1}{3!} t_{AWB}^{VIJ} (  E^{A\alpha W \beta B}_{V \alpha I\beta J} - I^W_V E^{A\alpha B}_{I\alpha J}      )  |\Phi \rangle,  \label{t3-eaeacc} \\
\frac{1}{3!} t_{AWX}^{VIJ} E^{AWX}_{VIJ}  |\Phi \rangle   &=  \frac{1}{3!} t_{AWX}^{VIJ} (  E^{A\alpha W\beta X\beta}_{V \alpha I\beta J\beta} - I^W_V E^{A\alpha X\beta}_{I\alpha J\beta} -   I^X_V E^{A\alpha W\beta}_{J\alpha I\beta}    )  |\Phi \rangle  \\
 \frac{1}{3!} t_{ABX}^{VWI} E^{ABX}_{VWI}  |\Phi \rangle &=  \frac{1}{3!} t_{ABX}^{VWI} (  E^{A\alpha B \alpha X \beta}_{V \alpha W\alpha I\beta} - I^X_W E^{A\alpha B\alpha}_{V\alpha I\alpha}  -  I^X_V E^{A\alpha B\alpha}_{I\alpha W\alpha}    )  |\Phi \rangle
\end{align}
provide forms in $\mu_3^{\textrm{PSA}[1|2|3]}$ of the spin-adapted excitation manifolds in Tables \ref{tab2}.
By comparing with the $T_2$ term on the reference ROHF state $|\Phi\rangle$, we obtain expressions with the $T_3$ spin adaptation in Table \ref{tab1}. For instance, if we revise the closed-external excitation second order excitation as
\begin{align}
\frac{1}{2} t_{AB}^{IJ} E_{IJ}^{AB} &\rightarrow \frac{1}{2} t_{A\alpha B}^{I\alpha J} E_{I\alpha J}^{A\alpha J} + \frac{1}{2} t_{A\beta B}^{I\beta J} E_{I\beta J}^{A\beta J} \\
 t_{A\alpha B}^{I\alpha J}  &=  t_{A B}^{I J} - \frac{2}{3!} t^{VIJ}_{AWB}I_V^W \label{rev-t2eecc} \\
 t_{A\beta B}^{I\beta J}  &=  t_{A B}^{I J}
\end{align}
where the second term in Eq. (\ref{rev-t2eecc}) will match the term with $I_V^Y$ in Eq. (\ref{t3-eaeacc}). As an alternative, the spin adaptation may be derived from generalized norm ordering as described in appendix \ref{appendix_c}.

The effects of $T_1 T_2 $ and $T_1^3$ on the reference state $|\Phi \rangle$ provide expressions of spin adapted amplitudes in Table \ref{tab1}. One subtlety is,
the exponential form of CC excitation, $e^T$, will generate non-linear excitations, e.g., $T_1^2$.
Once the single excitation operator with amplitude $t_A^I E_I^A$ was modified to
$ \left( t_{A}^{I} - \frac{1}{2}  t_{AV}^{WI} I_W^V \right) E_{I\alpha}^{A\alpha} +  t_{A}^{I} E_{I\beta}^{A\beta} $ by Eqs. (\ref{c2_1}) -(\ref{c2_3}),
the non-linear excitation will include additional terms from modification of excitation operators Eqs. (\ref{c2_1}) -(\ref{c2_3}). Take $t^{I\alpha}_{A\alpha}  t^{J\beta}_{B\beta}$ for instance,  under $[1|2]$ level of PSA, it will include $-\frac{1}{2} t_B^J  t_{AV}^{WI} I_W^V$. This term belongs to the spin adaptation of  $T_1 T_2 | \Phi \rangle$ in Table \ref{tab1}. In implementations, if both $T_2$ and $T_1 T_2$ spin adaptations in Table \ref{tab1} are included, the effect of the aforementioned additional term, $- \frac{1}{2} t_B^J t_{AV}^{WI} I_W^V$, needs to be excluded in the RHS of $T_1 T_2$ row in Table \ref{tab1} to avoid duplication. 

This subtlety does not invalid the effectiveness of modifying inactive excitations to match spin adaptation from fully spin-adapted expansion acting on the reference ROHF state, since the influence is subjected to and can be corrected at higher orders. For a system with a finite number of electrons, the excitation operator will be terminated by the number of electrons. The low-order excitation operators are suppressed by large denominators. For example, in $T_1^2/2!$ the denominator is 2. In $T_1 T_2/2 + T_2 T_1/2$, the combined denominator is 1. That suggests these spin adaptations may be effective.

Similarly,  modifications of fourth-order tensors on the reference state $|\Phi \rangle$:
\begin{align}
 \frac{1}{4!} t_{AWBC}^{VIJK} E^{AWBC}_{VIJK}  |\Phi \rangle &= \frac{1}{4!} t_{AWBC}^{VIJK} ( E^{A\alpha W\beta BC}_{V\alpha I\beta JK} - I^W_V  E^{A\alpha BC}_{I\alpha JK} ) |\Phi \rangle \\
\frac{1}{4!} t_{AWXB}^{VIJK} E^{AWXB}_{VIJK}  |\Phi \rangle &= \frac{1}{4!} t_{AWXB}^{VIJK} ( E^{A \alpha W \beta X \beta B}_{V\alpha I\beta J\beta K} - I^W_V E^{A\alpha X\beta B}_{I\alpha J\beta K} - I^X_V E^{A\alpha W \beta B }_{J\alpha I\beta K}  ) |\Phi \rangle  \\
\frac{1}{4!} t_{AWXY}^{VIJK} E^{AWXY}_{VIJK}  |\Phi \rangle &= \frac{1}{4!} t_{AWXY}^{VIJK} ( E^{A \alpha W \beta X \beta Y\beta}_{V\alpha I\beta J\beta K\beta} - I^W_V E^{A\alpha X\beta Y\beta}_{I\alpha J\beta K\beta} - I^X_V E^{A\alpha W \beta Y \beta}_{J\alpha I\beta K\beta}  \nonumber \\
& - I^Y_V E^{A\alpha W \beta X \beta}_{K\alpha I\beta J\beta}  ) |\Phi \rangle  \\
\frac{1}{4!} t_{ABXC}^{VWIJ} E^{ABXC}_{VWIJ}  |\Phi \rangle &= \frac{1}{4!} t_{ABXC}^{VWIJ} ( E^{A\alpha B\alpha X\beta C}_{V\alpha W\beta I\beta  J }
- I^X_W E^{A\alpha B\alpha C}_{V\alpha I\alpha J} - I^X_V E^{A\alpha B\alpha C}_{I\alpha W\alpha J} ) |\Phi \rangle  \\
\frac{1}{4!} t_{ABCY}^{VWXI} E_{VWXI}^{ABCY} |\Phi \rangle &= \frac{1}{4!} t_{ABCY}^{VWXI} ( E_{V\alpha W\alpha X\alpha I\beta}^{A\alpha B\alpha C\alpha Y\beta} - I^Y_X E^{A\alpha B\alpha C\alpha }_{V\alpha W\alpha I\alpha }  - I^Y_W E^{A\alpha B\alpha C\alpha}_{V\alpha I\alpha X\alpha} \nonumber \\
&- I^Y_V E^{A\alpha B\alpha C\alpha}_{I\alpha W\alpha X\alpha} ) |\Phi \rangle  \\
\frac{1}{4!} t_{ABXY}^{VWIJ} E_{VWIJ}^{ABXY} |\Phi \rangle &= \frac{1}{4!} t_{ABXY}^{VWIJ} (  E_{V\alpha W\alpha I\beta J\beta}^{A\alpha B\alpha X\beta Y\beta}
-  I^X_W E^{A \alpha B\alpha Y\beta }_{V\alpha I\alpha J\beta } - I^Y_W E^{A\alpha B\alpha X\beta}_{V\alpha J\alpha I\beta} \nonumber \\
& - I^X_V E^{A\alpha B\alpha Y\beta}_{I\alpha W\alpha J\beta} - I^Y_V E^{A\alpha B\alpha X\beta}_{J\alpha W\alpha I\beta}
+ I^X_V I^Y_W E^{A\alpha B\alpha}_{I\alpha J\alpha}
- I^X_W I^Y_V E^{A\alpha B\alpha }_{I\alpha J\alpha }  ) |\Phi \rangle
\end{align}
provide expressions of excitation manifolds $\mu_4^{\textrm{PSA}[1|2|3|4]}$ of the spin-adapted excitation manifolds in Table \ref{tab3}.

\section{Symmetries in PSA amplitude tensors}
\label{appendix_b}
Symmetries of tensorial quantities are crucial in the canonicalization procedure\cite{hirata2003tensor,Kallay:00,kallay2001higher, engels:11,wang2018simple,evangelista2022automatic}. In this appendix, we discuss the symmetries of amplitude tensors in the PSA framework.  When there is no unit matrix in the excitation operator associated with the amplitude tensor, symmetries can be derived by analogous with closed-shell spatial orbital and spin-orbital approaches. For instance, consider amplitude tensors with inactive indices
\begin{align}
t_{AB}^{IJ} E_{IJ}^{AB} |\Phi \rangle = t_{AB}^{IJ} E_{JI}^{BA} |\Phi \rangle =   t_{BA}^{JI} E_{IJ}^{AB} |\Phi \rangle \label{b1}
\end{align}
the first equality is based on the symmetry of excitation operator $E_{IJ}^{AB} = E_{JI}^{BA}$ \cite{Kutzelnigg:97} and the second equality is based on renaming dummy indices $AB \rightarrow BA $ and $IJ \rightarrow JI$.

Eq. \eqref{b1} leads to
\begin{align}
& ( t_{AB}^{IJ} - t_{BA}^{JI} ) E_{IJ}^{AB} |\Phi \rangle = 0 \\
\Rightarrow & t_{AB}^{IJ} = t_{BA}^{JI}
\end{align}
as it has been well-known \cite{Bartlett:09}.
 
For amplitude tensors with two active upper or lower indices,
\begin{align}
t_{AB}^{VW} E_{V\alpha W\alpha}^{A\alpha B\alpha} |\Phi \rangle = - t_{AB}^{VW} E_{W\alpha V\alpha}^{A\alpha B\alpha} |\Phi \rangle =  - t_{AB}^{WV} E_{V\alpha W\alpha}^{A\alpha B\alpha} |\Phi \rangle
\end{align}
leads to
\begin{align}
t_{AB}^{VW} = - t_{AB}^{WV}
\end{align}

When a unit matrix exists in the excitation operators, PSA amplitude tensors can still follow the symmetries as the closed-shell spatial orbital. For instance, the following derivations
\begin{align}
& \frac{1}{3!} t_{AWX}^{VIJ}  (  E^{A\alpha W\beta X\beta}_{V \alpha I\beta J\beta} - I^W_V E^{A\alpha X\beta}_{I\alpha J\beta} -   I^X_V E^{A\alpha W\beta}_{J\alpha I\beta}    )  |\Phi \rangle \label{eq_semi}  \\
&= \frac{1}{3!} t_{AXW}^{VJI} (  E^{A\alpha X\beta W\beta}_{V \alpha J\beta I\beta} - I^X_V E^{A\alpha W\beta}_{J\alpha I\beta} -   I^W_V E^{A\alpha X\beta}_{I\alpha J\beta}    )  |\Phi \rangle \\
&= \frac{1}{3!} t_{AXW}^{VJI} (  E^{A\alpha W\beta X\beta}_{V \alpha I\beta J\beta} - I^W_V E^{A\alpha X\beta}_{I\alpha J\beta} -   I^X_V E^{A\alpha W\beta}_{J\alpha I\beta}   )  |\Phi \rangle \label{b8}
\end{align}
The second line is obtained by renaming dummy indices $IJ$ and $WX$. The third line is obtained by the anticommuting relation between $W\beta$ and $X\beta$ and  between $I \alpha$  and $J \alpha $.  Eqs. \eqref{eq_semi} and \eqref{b8} lead to
\begin{align}
&\frac{1}{3!} ( t_{AWX}^{VIJ} -  t_{AXW}^{VJI}  )   (  E^{A\alpha W\beta X\beta}_{V \alpha I\beta J\beta} - I^W_V E^{A\alpha X\beta}_{I\alpha J\beta} -   I^X_V E^{A\alpha W\beta}_{J\alpha I\beta}    )  |\Phi \rangle = 0 \\
\Rightarrow & t_{AWX}^{VIJ} =  t_{AXW}^{VJI} \label{t_semi}
\end{align}

\section{Connections between PSA and generalized normal ordering}
\label{appendix_c}
In this appendix, we discuss connections between PSA and generalized normal ordering \cite{Kutzelnigg:97}, especially with presence of contractions between active indices. Notice that generalized normal ordering could remove the contraction of indices in a given excitation tensor \cite{mukherjee1979hierarchy, haque1984application,mukherjee1989use}, for instance,
\begin{align}
E_{VI}^{AW} |\Phi \rangle = ( \tilde{E}_{VI}^{AW} -  \gamma^{W\alpha}_{V\alpha} \tilde{a}^{I\alpha}_{A\alpha} ) |\Phi \rangle = ( \tilde{E}_{VI}^{AW} - \gamma^{W\alpha}_{V\alpha} a^{I\alpha}_{A\alpha}  ) |\Phi \rangle \label{appendix_c1}
\end{align}
as noticed by Herrmann and Hanrath in a different context, Eq. 11 in Ref. \cite{herrmann2022correctly}.

We thus rewrite the normal-ordered operator $ \tilde{E}_{VI}^{AW} $ acting on $|\Phi \rangle$ as
\begin{align}
\tilde{E}_{VI}^{AW} |\Phi \rangle &= ( E_{VI}^{AW} +\gamma^{W\alpha}_{V\alpha} a^{I\alpha}_{A\alpha}   ) |\Phi \rangle  \label{appendix_c2} \\
&=  E_{V \beta I\alpha}^{A\beta W\alpha} |\Phi \rangle  \label{appendix_c2-2}
\end{align}
The RHS of Eq. \eqref{appendix_c2} and the RHS of Eq. \eqref{appendix_c2-2} are essential the same as Eq.  \eqref{a3}. In this example,
$\tilde{E}_{VI}^{AW} |\Phi \rangle$ corresponds to the PSA[1] level of the spin adaptation as found in the beginning of PSA \cite{Janssen:91} and
$ \tilde{E}_{VI}^{AW} |\Phi \rangle  - \gamma^{W\alpha}_{V\alpha} a^{I\alpha}_{A\alpha}   | \Phi \rangle = \tilde{E}_{VI}^{AW} |\Phi \rangle  - \gamma^{W\alpha}_{V\alpha} \tilde{E}^{I\alpha}_{A\alpha}   | \Phi \rangle $
corresponds to the PSA$[1|2]$ level of the spin adaptation.

The above derivations imply one can start from a fully spin adapted form of open-shell CC, then perform a generalized normal ordering to derive the corrections of spin-adaptations, e.g., Table \ref{tab1}, in the PSA schemes than the direct derivation in appendix \ref{appendix_a}. One difference is, the modifications Eqs. (\ref{c2_1}) and (\ref{c2_2}) in $\left(T_1^{\mathrm{PSA}[1|2]}\right)^2$ introduces $- \frac{1}{2} t_B^J  t_{AV}^{WI} I_W^V  $, as mentioned in appendix \ref{appendix_a}. This term does not come from the contractions in fully spin-adapted $T_1 T_1$ nor $T_1 T_2$ acting on the reference ROHF state $\ket{\Phi}$. Further adjustment is needed to match the spin adaptations from the fully spin adapted expansion without double counting.

This connection suggests some similarities with the spin-free combinatoric approaches \cite{Datta:08, datta2011state, Datta:13, datta2014analytic, datta2015communication, datta2019accurate}. One difference is: in the spin-free combinatoric approaches, one starts from the fully spin-adapted excitation operators and performs normal orderings to avoid the $TT$ contractions. Nevertheless the fully spin-adapted $E_I^W$ and $E_V^A$ operators are still used. This would lead to the difference in $e^{-T}$ \cite{Janssen:91}. Therefore, the $TH$ contractions are presented in the linked form of these approaches \cite{datta2014analytic,datta2019accurate}. In any order of PSA schemes of the present work, no $TH$ contractions would appear.

\bibliographystyle{pccp}
\bibliography{topic_os}

\end{document}